\documentclass[preprint,superscriptaddress,onecolumn]{revtex4-1}
\usepackage{amsmath,graphicx}
\usepackage{siunitx}
\usepackage{dirtytalk}
\usepackage{microtype}
\usepackage{comment}
\usepackage{color}

\begin{document}

\title{Measuring picometre-level displacements using speckle patterns\\ produced by an integrating sphere}

\author{Morgan Facchin} \email{mf225@st-andrews.ac.uk}
\affiliation{SUPA, School of Physics and Astronomy, University of St Andrews, North Haugh, St Andrews KY16 9SS, UK}

\author{Graham D. Bruce}
\affiliation{SUPA, School of Physics and Astronomy, University of St Andrews, North Haugh, St Andrews KY16 9SS, UK}

\author{Kishan Dholakia}
\affiliation{SUPA, School of Physics and Astronomy, University of St Andrews, North Haugh, St Andrews KY16 9SS, UK}
\affiliation{Department of Physics, College of Science, Yonsei University, Seoul 03722, South Korea}
\affiliation{School of Biological Sciences, The University of Adelaide, Adelaide, South Australia, Australia}

\begin{abstract}
As the fields of optical microscopy, semiconductor technology and fundamental science increasingly aim for precision at or below the nanoscale, there is a burgeoning demand for sub-nanometric displacement and position sensing. We show that the speckle patterns produced by multiple reflections of light inside an integrating sphere provide an exceptionally sensitive probe of displacement. We use an integrating sphere split into two equal and independent hemispheres, one of which is free to move in any given direction. The relative motion of the two hemispheres produces a change in the speckle pattern from which we can analytically infer the amplitude of the displacement. The method allows displacement measurement with uncertainty as small as 40 pm ($\lambda/20,000$) in a facile implementation. We show how, under realistic experimental parameters, the uncertainty in displacement could be improved to tens of femtometres, or $\lambda/10^{7}$.
\end{abstract}
\maketitle

\section{Introduction}
The ability to measure sub-nanometre displacements is increasingly vital to a wide range of fields of science and technology. State-of-the-art gravimeters and seismometers seek to measure sub-nanometre changes in the position of oscillators \cite{middlemiss2016measurement}. As the semiconductor industry moves inexorably to smaller scales, the demands on the accurate positioning of components grows, and already requires sub-nanometre precision \cite{lee2011wafer}. A related problem is found in super-resolution microscopy. While techniques such as PALM \cite{betzig2006imaging} and STORM \cite{rust2006stochastic} allow imaging with nanometre-resolution, the long acquisition times require the samples under investigation to drift by significantly smaller distances over the timescale of the experiments. Progress in super-resolution microscopy therefore requires simultaneous improvement of feedback-controlled sample positioning with sub-nanometre stability to avoid motion-blur.

A number of methods have emerged to resolve displacement at or below the nanometre level \cite{de2019review,berkovic2012optical}. For example, tracking the light scattered by a point emitter \cite{tischler2018all} or a nano-antenna \cite{neugebauer2016polarization} can resolve displacements at sub-nanometre level. Phase variations of the superoscillatory fields generated by a Pancharatnam-Berry metasurface can resolve displacements of the metasurface at the nanometre-level \cite{Yuan2019ruler},
while the transverse displacement of two such phase gratings can be detected by the polarisation rotation this encodes on a laser beam, with 0.4 nm resolution  (i.e. $\lambda/1600$, where $\lambda$ is the wavelength of the light) demonstrated in \cite{barboza2021ultra}. 

Going beyond traditional one-dimensional interferometers, a powerful tool for optical metrology can be provided by the two-dimensional random interference patterns known as speckle. Such speckle patterns are produced when light interacts with a disordered medium. They typically allow for facile, inexpensive setups, and have recently been applied to various topics in metrology including: the measurement of the polarisation \cite{kohlgraf2009spatially,facchin2020pol}, topological charge \cite{Mazilu12} and wavelength or spectrum of light \cite{Metzger17,Cao17,bruce19}; the refractive index \cite{Tran20,Cabral2020,Facchin2021ref} or temperature \cite{Cabral2020} of the environment; or the strain on the disordered medium \cite{Murray19}. 

On the particular topic of displacement, earlier works made use of speckle to measure in-plane displacement by autocorrelation \cite{archbold70} and spatial mapping of displacement \cite{leendertz1970}. Many improvements and variations of these have since been made \cite{yang2014review}. 
The current most accurate speckle-based method for displacement measurement, to our knowledge, tracks singularities in the pseudo phase of a moving speckle pattern to reach nanometric precision \cite{wang2006core,wang2006vortex}. 
These methods rely on the ability to directly image the translation of features of interest, in a regime where the speckle pattern translates along the direction of displacement of the scattering medium. We suggest a fundamentally different and more sensitive approach, which is to find a geometry where small displacements induce a large morphological change in the speckle pattern. This obviates the need for any magnification system to resolve translational motion.

Although speckle can be generated in many ways, we have recently analytically shown that an integrating sphere offers an optimal scattering medium, due to the large spread of optical path lengths inherent in this geometry \cite{Facchin_model}. In the context of measuring the wavelength of monochromatic light, the speckle from a typical integrating sphere is four orders of magnitude more sensitive to wavelength changes than the speckle produced in more-ubiquitous multi-mode fibre \cite{Facchin_model}. The integrating sphere has also been used recently to advance the speckle-based measurement of changes in refractive index of gases at the level of one part in $10^{9}$ \cite{Facchin2021ref}.

Here, we show that the speckle from an integrating sphere also provides a highly sensitive probe of displacement. In contrast to previous usage, we use a novel integrating sphere geometry comprising of two unattached hemispheres. We perform displacement measurements in both axial and transverse directions with an uncertainty of tens of picometres, that is $\lambda/20,000$ where $\lambda$ is the wavelength of light used to generate the speckle. The method is based on the evaluation of the speckle similarity, that we relate to displacement following an analytical model. We show how the measurement uncertainty could be reduced to tens of femtometres ($\lambda/10^{7}$) under realistic optimisation of the experimental parameters. We also demonstrate a method utilising a virtual hemisphere to perform strictly one-dimensional displacement measurements.

\section{Similarity Profile}
We consider an integrating sphere split into two independent hemispheres, one of which is able to move by an amount $x$ in any given direction. 
In order to infer the displacement from the speckle change, we first need a tool to quantitatively measure that change. This is done using the speckle similarity, or correlation, which is given by 
\begin{equation} \label{eq:correl}
S= \Big\langle  \Big( \frac{I_{i} - \langle I_{i}\rangle}{\sigma} \Big)\Big( \frac{I_{i}'-\langle I_{i}'\rangle}{\sigma'} \Big) \Big\rangle ,
\end{equation} 
with $I$ and $I'$ two speckle images, $\sigma_{I}$ and $\sigma_{I'}$ their respective standard deviation, and the angular brackets denote averaging over the index $i$, or the image. This gives a value of~1 for identical images, and decreasing values as they diverge from one another.

In the case of speckle patterns produced by multiple reflections of light in an integrating sphere, it was shown in \cite{Facchin_model} that for two speckle patterns taken before and after some generic transformation, the similarity is given by 
\begin{equation} \label{Sgeneral}
S =   \frac{ 1 }
{\big (1-\frac{\sigma^2}{2\ln{\rho}} \big)^2+\big (\frac{\mu}{\ln{\rho}} \big)^2     },
\end{equation}

\noindent where $\mu$ and $\sigma^2$ are respectively the average and variance of the phase shift induced by the transformation on a single pass of light through the sphere, with $\rho$ the sphere's surface reflectivity. A single pass (or chord) is defined as a straight line joining two points of the sphere.


In our case, the phase shift is related to the change in length of the chords resulting from the displacement. We derive analytically the expressions of $\mu$ and $\sigma^2$ (see Appendix A) in two particular cases, when the displacement is along the symmetry axis (denoted as axial motion), and perpendicular to the symmetry axis (denoted as transverse motion). The symmetry axis and both directions are shown in Fig. \ref{fig:setup}.

In the axial case, we find $\mu=kx/3$ and $\sigma=kx\sqrt{5}/6$. Inserting this in (\ref{Sgeneral}), we can neglect the $\sigma$ term and obtain   
\begin{equation} \label{eq:Sax}
S=  \frac{1}{1+\left(\frac{kx}{ 3\ln{\rho}}\right)^2 },
\end{equation} 
which is a Lorentzian with an HWHM (half width at half maximum) given by $3\lambda \left | \ln{\rho}\right |/2\pi  \approx  0.5\lambda\left | \ln{\rho}\right | $.

In the transverse case, we have $\mu=0$ and $\sigma=kx/\sqrt{8}$, leading to
\begin{equation} \label{eq:Str}
S=  \frac{1}{\left (1-\frac{\left(kx\right)^2}{16\ln{\rho}} \right)^2 },
\end{equation} 
which is the square of a Lorentzian with an HWHM given by $\sqrt{16(\sqrt{2}-1)}\lambda \sqrt{\left | \ln{\rho}\right |}/2\pi  \approx  0.4\lambda\sqrt{\left | \ln{\rho}\right | }$.

The axial motion imparts a greater change to the speckle pattern than the transverse one. This can be understood qualitatively, as in the axial case the chords increase in length in average, resulting in a systematic phase increase after each single pass. In the transverse case however, some chords can compensate each other, leading to smaller change. 

For typical parameters such as $\lambda=780$~nm and $\rho=0.9$, the HWHM is 39 nm in the axial case, and 104 nm in the transverse case. We note that the HWHM is not the resolution limit of the approach - the ultimate resolution of the method depends on the smallest detectable variation in similarity (see Section \ref{sect:unc}). 
Note that the sensitivity of the speckle pattern to displacement is independent of the size of the sphere, contrary to what is found for wavelength and refractive index variation \cite{Facchin_model}. 

The knowledge of the similarity profile allows one to infer the displacement between two given instants, by simply applying the reciprocal function of the appropriate profile and using the value of the similarity of the two corresponding speckles.

\section{Experimental Implementation}
\begin{figure}[h!] 
\centering\includegraphics[width=14cm]{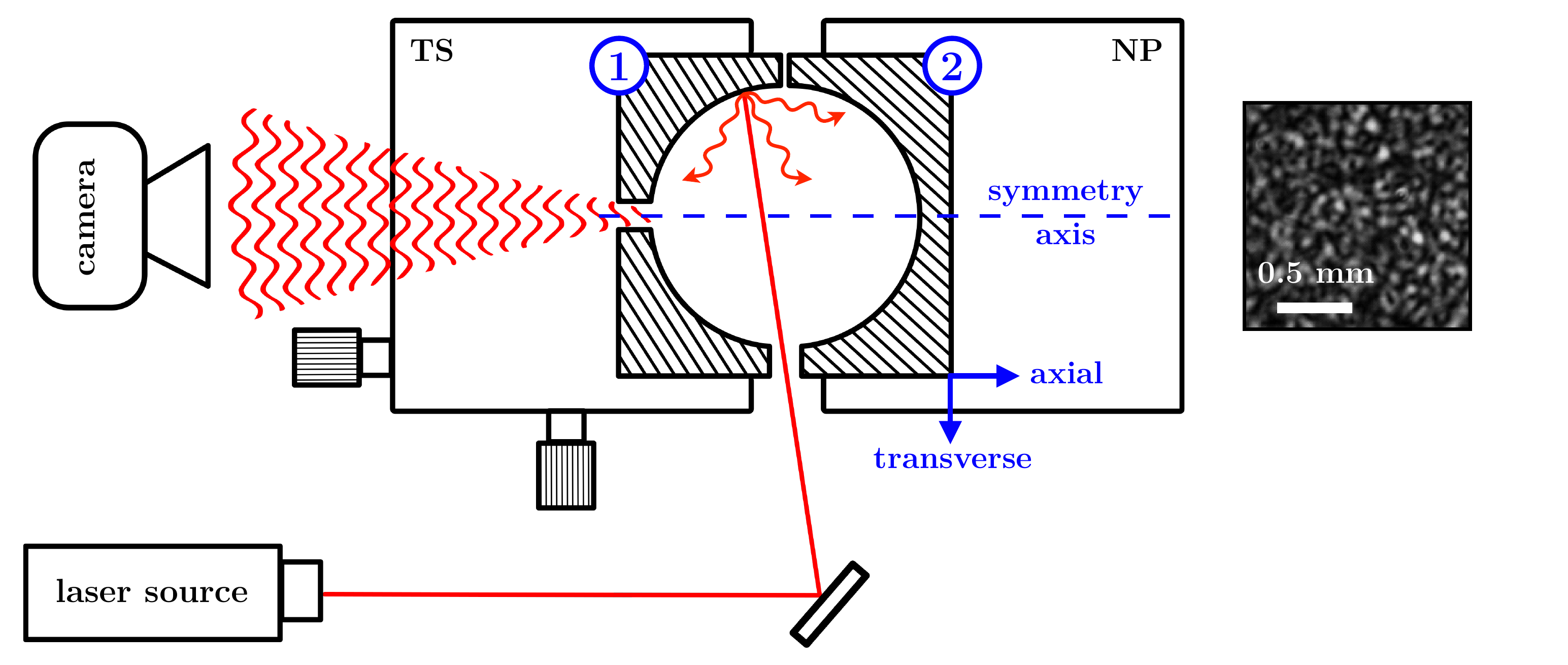}
\caption{Experimental setup. Laser light enters the integrating sphere and produces a speckle pattern recorded on a camera. Hemisphere (1) rests atop a manual 3D translation stage (TS) for coarse alignment with hemisphere (2), which rests on a motorised nanopositioner (NP). The latter is moved at a constant speed while the changing speckle pattern is recorded. The symmetry axis, the axial and transverse directions, and an example of speckle pattern are shown. }
\label{fig:setup}
\end{figure}

\begin{figure}[h!] 
\centering\includegraphics[width=11cm]{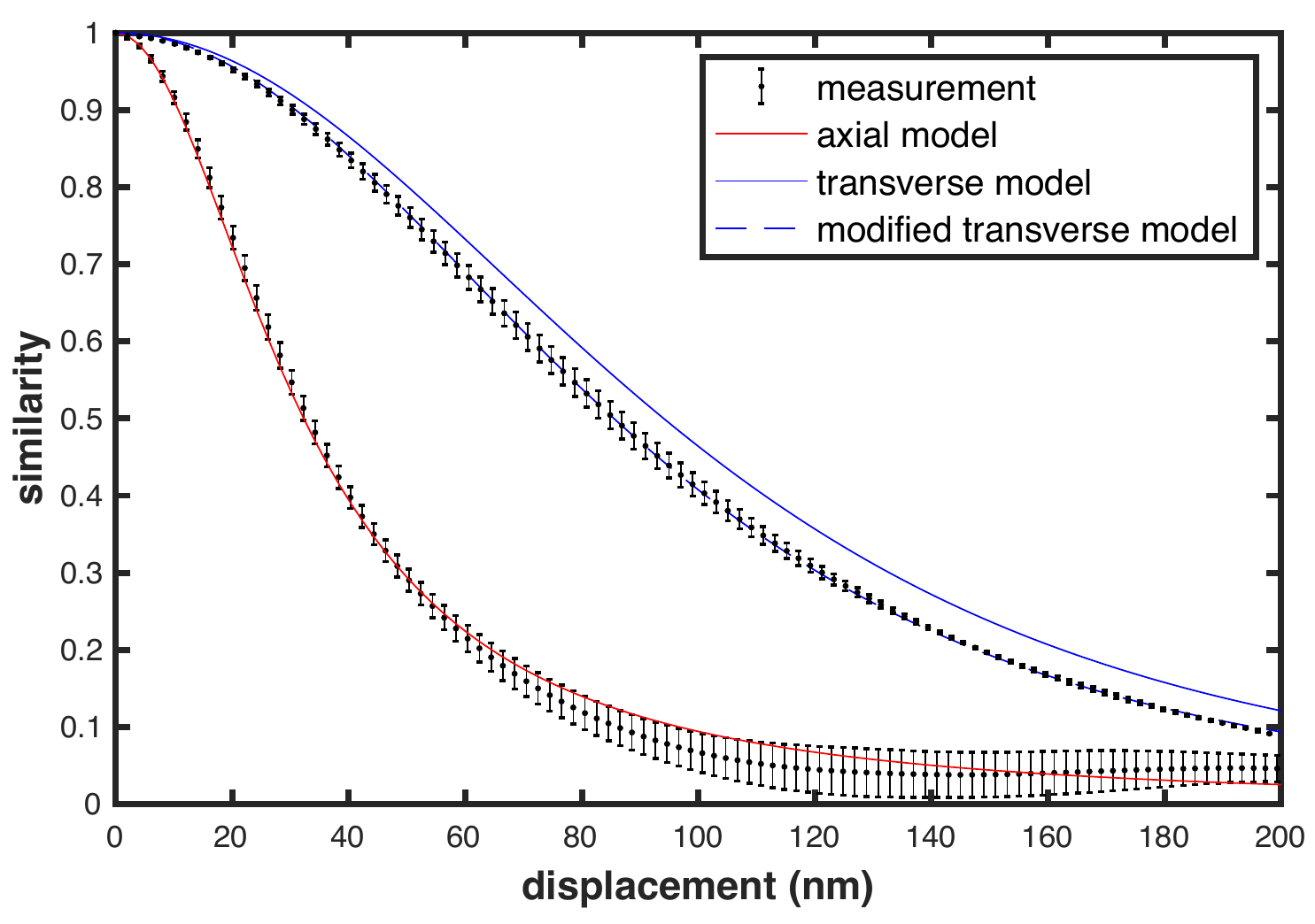}
\caption{Speckle similarity profiles as a function of displacement. Experimental data (black), axial theoretical profile (red), and transverse theoretical profile (blue). The HWHM is 32 nm in the axial case and 85 nm in the transverse case. The observed transverse profile shows a small systematic deviation from the model, which can be explained by an imperfect Lambertian reflectance and/or a non uniform reflectivity. This is taken into account in the modified model (dashed blue). 
}
\label{fig:disp}
\end{figure}

In this section we experimentally verify the relations (\ref{eq:Sax}) and (\ref{eq:Str}) with the setup described in Fig. \ref{fig:setup}. A laser beam of 780 nm wavelength, 10 mW power, and a coherence length of a few kilometres (Toptica DLPro) is injected into an integrating sphere, and the resulting speckle pattern in collected on a CMOS camera (Mikrotron MotionBLITZ EoSens mini2). We use $200\! \times \! 200$-pixels images which offers a good compromise between computation time and variance of the computed similarity profile. The light enters and escapes the sphere via two 3 mm diameter holes. We use a 1.25 cm radius sphere, carved in a 3 cm edge aluminium cube and coated with Spectraflect to give a near Lambertian reflectance with reflectivity $\rho=0.918\pm0.008$. The sphere is made of a fixed and a moving half. The fixed half rests on a manual translation stage, for coarse alignment of the two halves, and the moving half rests on a 3D precision stage capable of nanometer precision (PI P-733.3DD). We displace the moving hemisphere in the axial and transverse (horizontal) direction at a constant speed of $0.1$~\si{\micro\meter\per\second} while the changing speckle is recorded. We then extract the similarity profiles by applying expression (\ref{eq:correl}) between one reference and subsequent images. By choosing several different reference images we extract different similarity profiles, which allows us to find the average profile with its corresponding standard deviation that we display as an error bar. The profiles found in this way are shown in Fig. \ref{fig:disp}, where the theoretical profiles (\ref{eq:Sax}) and (\ref{eq:Str}) are also shown. 

We find a good agreement for the axial profile, while the transverse profile shows a small deviation. This difference was found to be systematic and independent of many experimental parameters. In Appendix B, we show how this can be explained by a small deviation from the assumptions of our model, in particular the Lambertian reflectance and the uniformity of the reflectivity across the inner surface of the integrating sphere.

\section{Uncertainty} \label{sect:unc}

For measuring displacements much smaller than the HWHM of the similarity profile, we must notice that taking directly the reciprocal function of the similarity profile is not ideal, as we have a horizontal tangent at zero, meaning that a small displacement only imparts a small change to the value of the similarity. The optimal configuration to measure small displacement is to first apply an initial offset that brings the similarity to a value of 0.5, which is an inflection point in the case of a Lorentzian profile. As this is the point of maximum slope, the similarity is maximally sensitive to subsequent displacements. In other words, if the similarity is taken between two speckles before and after a displacement $x=x_0+\delta x$, with $x_0$ the HWHM, it will vary around a value of 0.5 with maximal sensitivity to $\delta x$. Applying the reciprocal function of (\ref{eq:Sax}), namely $x=x_0\sqrt{1/S-1}$, and subtracting the offset, gives $\delta x$.

\begin{figure}[h!] 
\centering\includegraphics[width=13cm]{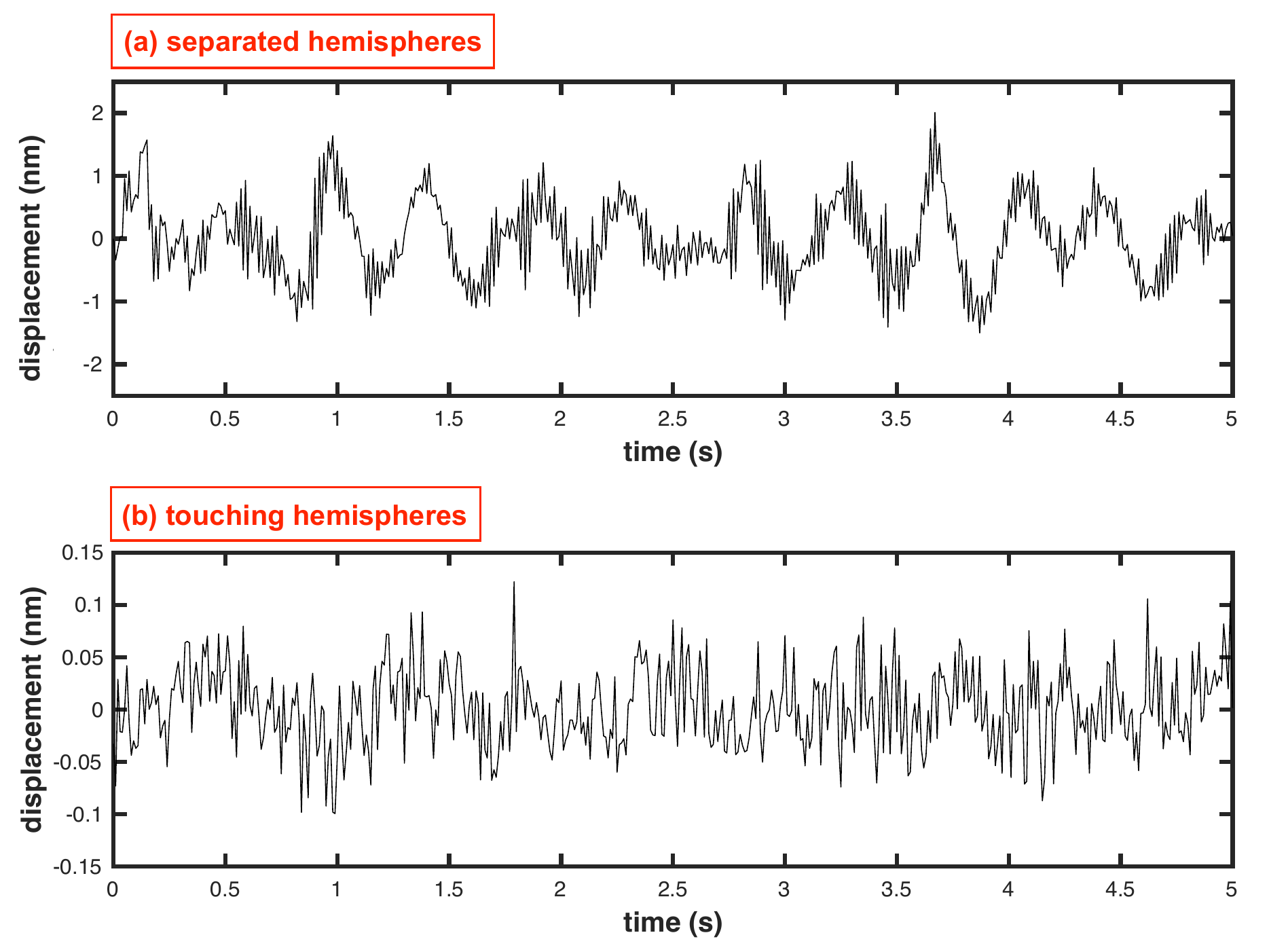}

\caption{Measured displacement as a function of time when the hemispheres are (a) separated and (b) touching. In (a) background mechanical perturbations are visible, with a standard deviation of 0.6 nm. In (b) any displacement is suppressed, leaving only measurement noise and other possible sources of speckle change. The standard deviation in that case is 40 pm, which defines our measurement uncertainty. }
\label{fig:noise}
\end{figure}

We want to determine the uncertainty on displacement in this optimal configuration. We implement this procedure in the following way. We start by applying the initial offset, which is done via a wavelength variation of 0.5 pm, instead of a displacement. This is physically equivalent to moving the hemisphere by an amount equal to the HWHM (32 nm). Indeed, any transformation that is dominated by the $\mu$ term in (\ref{Sgeneral}) can be used interchangeably to produce the same change in the speckle pattern (thermal expansion and refractive index variation are other examples \cite{Facchin2021ref}). A wavelength change instead of an actual displacement has the advantage of being applicable independently of the variable of interest, and can be done in a simple way by changing the current of the laser diode. We then compute the similarity between the speckle before and after the wavelength variation. Applying the reciprocal function of (\ref{eq:Sax}), namely $x=x_0\sqrt{1/S-1}$, and subtracting the offset gives the displacement happening after the wavelength variation. Finally, the signal is treated using a highpass Fourier filter, with a cutoff frequency of 1 Hz to remove any low frequency drift due to heating effects \cite{Facchin2021ref} (highpass function of Matlab). We perform this measurement in two cases, firstly with a gap left between the hemispheres of about 0.5 mm, and secondly with the hemispheres in direct contact. We show 5 seconds of such measurements in Fig. \ref{fig:noise} at a sampling frequency of 100 Hz. The measurements with separated hemispheres reveal mechanical perturbations with a standard deviation of 0.6 nm and a typical period of 0.4 s, which most likely originate from vibrations of the optical table or the stage. With the hemispheres in direct contact, any mechanical perturbations are greatly diminished, leaving only the measurement noise and other possible sources of speckle change (such as wavelength or temperature variations). The standard deviation of the signal in that case is 40 pm, which defines the uncertainty on the displacement measurement. 

The uncertainty can be expressed in more general terms as a function of the experimental parameters. At the inflection point, the uncertainty on displacement is given by the uncertainty on the similarity, divided by the slope of the similarity profile at that point. The uncertainty on the similarity is empirically found to be $\approx 0.1/\sqrt{N}$, with $N$ the number of pixels in the image. The slope of the similarity curve at the inflection point is $1/(2x_0)$, with $x_0=3\lambda\ln{\rho}/2\pi$ in the axial case. It follows that the uncertainty on displacement is approximately given by 
\begin{equation} \label{eq:uncert}
\Delta x \approx \frac{3\lambda\left | \ln{\rho}\right |}{10\pi\sqrt{N}}.
\end{equation} 

Currently, this uncertainty is principally limited by the reflectivity of our integrating sphere. Cone, et al, have demonstrated diffusely reflecting surfaces with reflectivity up to 0.99919 at 532 nm \cite{Cone15}. For an integrating sphere with such a level of reflectivity, illuminated with 532 nm light, equation (\ref{eq:uncert}) predicts a reduced uncertainty in the axial motion at the level of 200 fm, i.e. below $\lambda/10^{6}$. Further gains in uncertainty, at the cost of acquisition speed and data size, could be made by acquiring larger images. Using the full $2336\times1728$ pixels of our camera to acquire the speckle generated with one of these higher-reflectivity integrating spheres should allow the measurement of axial displacement of the hemispheres at the level of 20 fm, i.e. smaller than $\lambda/10^{7}$.

\section{Variation with a Virtual Hemisphere}
\begin{figure}[h!] 
\centering\includegraphics[width=11cm]{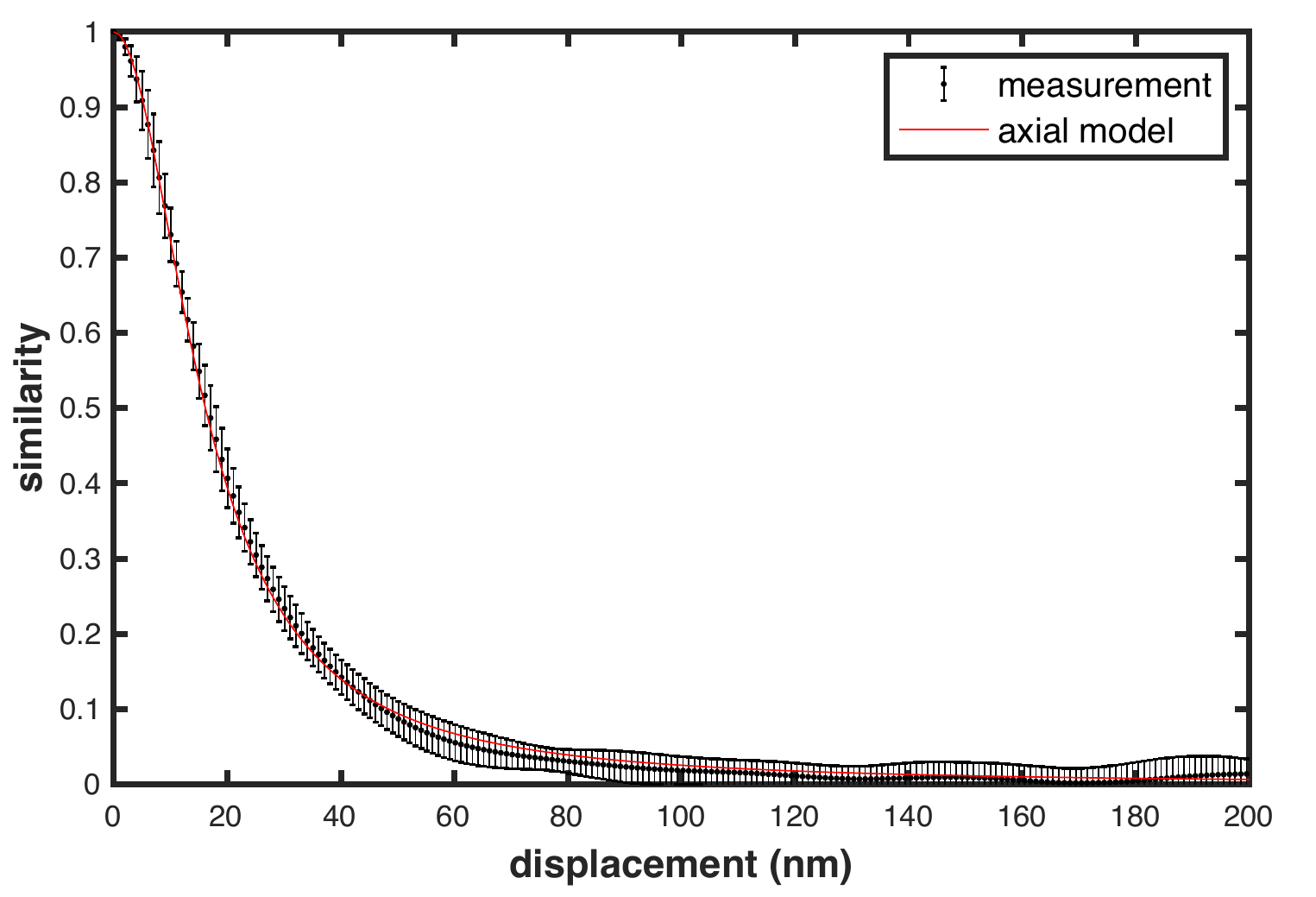}
\caption{Axial similarity profile in a modified version of the experimental setup, where the moving hemisphere is replaced by a flat mirror. The profile is also Lorentzian, with a HWHM of 16~nm, precisely half that found in the previous experiment. This finds a simple explanation in terms of the moving virtual image of the hemisphere. Here the axial model is obtained by multiplying $x$ by 2 in equation (\ref{eq:Sax}).}
\label{fig:mirror}
\end{figure}

We apply a small modification to the previous experimental setup, where the moving hemisphere is replaced by a flat mirror. In this arrangement, we find that the speckle becomes more sensitive to axial motion and insensitive to transverse motion. The similarity profile obtained for the axial motion is shown in Fig. (\ref{fig:mirror}). 

Although the geometry of the problem seems very different from the previous experiment, we find again a Lorentzian profile, with a HWHM precisely twice as small (the same factor of two was found with other spheres of different reflectivities and radii). This curious result can be understood in a simple way if we think of the image of the fixed hemisphere. The real and the virtual hemispheres form together a complete sphere, in which we can apply our previous model as if the whole sphere was real. Indeed, the motion of the mirror implies an axial motion of the virtual hemisphere, just like we had in the previous setup. The virtual hemisphere however, by its very nature, moves twice as much as the mirror, which explains the factor of two. Also, transverse motion does not change the position of the virtual hemisphere, which is consistent with the absence of speckle change in that case. 

In this interpretation we neglect the effect of the mirror's own reflectivity ($>$0.99 measured at 6° and 45° angle of incidence), as it is much higher than that of the sphere (0.918). If both reflectivities were closer together, the result would be less trivial. In fact, as any path reaching the virtual hemisphere has to undergo a reflection on the mirror, the similarity profile would be that of a real sphere divided in two regions of different reflectivities, one $\rho$ and one $\rho'\rho$ (with $\rho'$ the mirror's reflectivity), which would surely result in a similarity profile implying a non-trivial combination of $\rho$ and $\rho'$. 

We note in passing that this arrangement allows one to filter out any transverse component of the mirror's motion, and retain only the axial component, which may be beneficial in applications where detection of only one direction of motion is desired.

\section{Summary and conclusion}

In this work, we used the speckle pattern produced by two unattached hemispheres to measure displacements with an uncertainty of 40 pm ($\lambda/20,000$) in the axial direction, or below the size of a hydrogen atom. In future work, the use of a higher reflectivity coating on the hemispheres and a larger image array could be used to probe displacements on the level of tens of femtometres ($\lambda/10^{7}$). We demonstrated that, by replacing one hemisphere with a highly-reflective mirror, the displacement sensitivity is enhanced by a factor two in the axial direction while vanishing in the transverse direction, which gives an elegant method to selectively probe one direction of motion. 

The method presented here has the advantage of a very simple implementation, as it only involves a single beam with no special alignment needed: the two hemispheres only require a coarse alignment. Moreover, as the displacement is captured by a morphological change in the speckle as opposed to a small translation of a feature of interest, the approach does not require expensive magnification systems, which is in contrast to previous approaches \cite{Yuan2019ruler,wang2006core}.

We note that, interestingly, sensitivity is independent of the radius of the sphere, indicating that similar performance could be achieved in a more compact fashion. However the model used in this work would remain valid only for a sphere radius large enough compared to the wavelength. Realising measurements of such small displacements in isolation will require the use of ultra-stable optomechanics, but could be applied to accurate displacement measurement in fields ranging from force sensing to microscope sample positioning or seismology.


\appendix
\section{Derivation of Similarity Profiles} \label{sec:sim}
We derive here the expression of the similarity profiles for the axial and transverse motion. We first want to express the phase shift on a chord resulting from the displacement of the hemisphere (a chord is a straight line joining two points of the sphere). The effect of the displacement is to change the lengths of the chords, and therefore the length of propagation of light, which results in a phase shift. Along a chord of length $z$, light acquires a phase $kz$, with $k$ the wavenumber. When the hemisphere is displaced, this phase varies by $k\Delta z$, with $\Delta z$ the change in length. For small displacements, $\Delta z$ is given by $\boldsymbol{u}\cdot\boldsymbol{x}$, with $\boldsymbol{u}$ a unit vector parallel to the chord (oriented from the fixed to the moving point), and $\boldsymbol{x}$ the displacement vector of the moving hemisphere. Moreover, no phase shift occurs if the chord starts and ends on the same hemisphere. This can be modelled by a variable $s$, that takes a value of 0 when both ends of the chord belong to the same hemisphere, and 1 otherwise. It follows that the phase shift induced by the displacement on a chord can be expressed as
\begin{equation}
    \phi=k \, \boldsymbol{u}\cdot\boldsymbol{x} \, s.
\end{equation}
The terms $\mu$ and $\sigma^2$ in (\ref{Sgeneral}) are the mean and variance of $\phi$.
We therefore seek to express the following quantities 
\begin{equation} 
\mu=\overline{\phi} \quad \quad \quad \sigma^2=\overline{\phi^2}-\overline{\phi}^2,
\end{equation}
where the bar indicates averaging over random chords in the sphere. A random chord is a chord whose endpoints are chosen with a uniform probability distribution across the inner surface.

Let us express $\phi$ in more explicit terms before computing the averages. We express $\boldsymbol{u}$ and $\boldsymbol{x}$ in a spherical coordinate system, whose origin is at the centre of the sphere and $z$ axis confounded with the symmetry axis, with the moving hemisphere being on the positive side. In this system we use the spherical angles $\theta\in[0 \; \pi/2]$ and $\varphi\in[0 \; 2\pi]$. The restricted range of $\theta$ is chosen to guarantee that $\boldsymbol{u}$ is uniquely defined for a given chord, and points from the fixed to the moving hemisphere. In the axial case, we have $\boldsymbol{x}=x\boldsymbol{\hat{z}}$, and therefore $\phi=k x \cos{\theta} \, s$, with $\theta$ the angle between the chord and the symmetry axis. With this we can express $\mu$ as
\begin{equation} 
\mu=\iint  k x \cos{\theta} \, s \, f(\theta,s)\text{d}\theta\text{d}s,
\end{equation}
with $f(\theta,s)$ the joint probability distribution of $\theta$ and $s$. As $s$ is a discrete variable, this can be recast as 
\begin{equation} 
\mu=\int_{0}^{\pi/2} k x \cos{\theta} \, s \, P(s\!=\!1|\theta)f(\theta)\text{d}\theta,
\end{equation}
with $P(s\!=\!1|\theta)$ the probability of $s$ being 1 for a given $\theta$, and $f(\theta)$ the probability distribution of $\theta$. 

We can find $f(\theta)$ in the following way. By symmetry, the distribution of random chords is isotropic. Therefore, the set of all possible $\boldsymbol{u}$ vectors forms a uniform unit hemisphere. The number of chords contained around a certain $\theta$ is then proportional to the surface element in our spherical system, which is proportional to $\sin{\theta}$. After normalisation, we simply have $f(\theta)=\sin{\theta}$. 

\begin{figure}[h!] 
\centering\includegraphics[width=12cm]{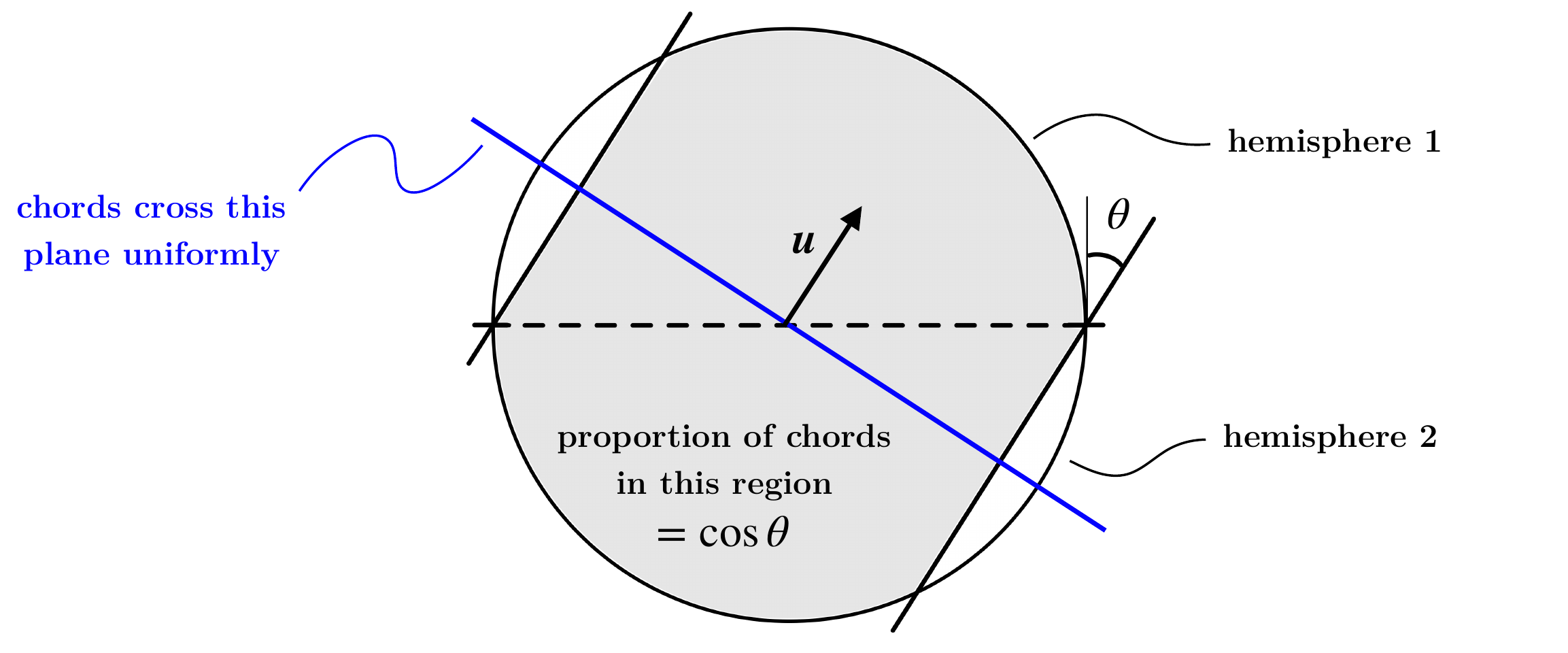}
\caption{Side view with the two hemispheres separated by the dashed line. The chords contained in an infinitesimal solid angle around the direction $\boldsymbol{u}$ cross the blue plane with a uniform density. It follows that the proportion of those chords that touch both hemispheres is $\cos{\theta}$.}
\label{fig:sphere}
\end{figure}

Finding $P(s\!=\!1|\theta)$ is more subtle. Let us consider the set of chords contained in an infinitesimal solid angle around the direction $\boldsymbol{u}$ forming an angle $\theta$ with the symmetry axis. It can be shown that those chords cross any plane perpendicular to $\boldsymbol{u}$ with a uniform density. As $P(s\!=\!1|\theta)$ corresponds to the proportion of those chords that touch both hemispheres, we find by geometry $P(s\!=\!1|\theta)=\cos{\theta}$ (see Fig. \ref{fig:sphere}). 


We have now everything in hand to to compute the integrals: 
\begin{equation} \label{mu_ax}
\begin{split}
\mu&=\int_{0}^{\pi/2} k x \cos{\theta} \cos{\theta}  \sin{\theta}\text{d}\theta\\
\mu&=\frac{k x}{3}.
\end{split}
\end{equation}

Similarly for $\sigma$: 
\begin{equation} \label{sig_ax}
\begin{split}
\sigma^2&=\int_{0}^{\pi/2} (k x \cos{\theta})^2 \cos{\theta} \sin{\theta}\text{d}\theta - \mu^2\\
\sigma^2&=\frac{(k x )^2}{4}-\frac{(k x )^2}{9}=\frac{5}{36}(k x )^2.
\end{split}
\end{equation}

In the transverse case, we have $\boldsymbol{x}=x\boldsymbol{\hat{x}}$ (or any direction in the $xy$ plane), and therefore $\phi=k x \sin{\theta}\cos{\varphi} \, s$, with $\theta$ the angle between the chord and the symmetry axis, and $\varphi$ the azimuthal angle of the chord with respect to the $\hat{x}$ direction. This invokes the distribution of the azimuthal angle $f(\varphi)$, which by symmetry is uniform, giving $f(\varphi)=1/2\pi$. This leads to
\begin{equation} \label{mu_tr}
\begin{split}
\mu&=\int_{0}^{2\pi}\int_{0}^{\pi/2} k x \sin{\theta}\cos{\varphi} \cos{\theta}  \sin{\theta} \frac{1}{2\pi}\text{d}\theta\text{d}\varphi\\
\mu&=0
\end{split}
\end{equation}

\begin{equation} \label{sig_tr}
\begin{split}
\sigma^2&=\int_{0}^{2\pi}\int_{0}^{\pi/2} (k x \sin{\theta}\cos{\varphi} )^2 \cos{\theta} \sin{\theta} \frac{1}{2\pi}\text{d}\theta\text{d}\varphi\\
\sigma^2&=\frac{(k x )^2}{8}.
\end{split}
\end{equation}

We can also perform the calculation for an arbitrary direction of displacement forming an angle $\beta$ with the symmetry axis, in which case we find 
\begin{equation} 
\begin{split}
\mu&=\frac{k x }{3}\cos^2{\beta}\\
\sigma^2&=(k x )^2\left( \frac{1}{8}\sin^2{\beta} + \frac{5}{36}\cos^2{\beta} \right),
\end{split}
\end{equation}
yielding the axial and transverse results from $\beta=0$ and  $\beta=\pi/2$ respectively.

Results (\ref{mu_ax}-\ref{sig_tr}) were verified numerically by computing length variations of random chords in a sphere, modelled by an assembly of uniformly distributed random points across the unit sphere, after translating half of it in the axial and transverse direction. We found very good agreement to 1 part in 1000.

\section{Modified Model} \label{sec:mod}
Here we show the effect of a deviation from either the Lambertian reflectance or the uniform reflectivity on the similarity profile. Following the derivation given in \cite{Facchin_model}, the general expression of the similarity (\ref{Sgeneral}) comes from the integral
\begin{equation} \label{integral}
\begin{split}
S&= \left | \int_0^{\infty} -\ln\!\rho\;\left(\rho  \overline{e^{i\phi}} \right)^N dN\right |^2,
\end{split}
\end{equation}
where $\phi$, the phase shift applied to the light on a chord, is Gaussian with mean $\mu$ and variance $\sigma^2$, and the overline designates averaging over random chords. 

The term $\rho  \overline{e^{i\phi}}$ in turn comes from the expression $\sum{Te^{i\phi}}$, appearing in a discrete model of the system where the inner surface is divided into $M$ discrete elements (the sum is over $M$ chords, and $T$ is the power loss after one reflection). Assuming a Lambertian reflectance, uniform reflectivity, an spherical geometry, $T$ is simply $\rho/M$, leading to $\rho \overline{e^{i\phi}}$. 

We can describe any deviation from the Lambertian reflectance or the uniform reflectivity by a dimensionless function $g$, which we include as a factor of $T$. This function can contain a direction dependence to model an excess of power in the specular direction for example, or a position dependence to model a non uniform reflectivity. We have $g=1$ for a Lambertian reflectance and uniform reflectivity, and $\overline{g}=1$ in any case, by conservation of power. With this definition, the previous $\rho \overline{e^{i\phi}}$ term becomes $\rho \overline{g e^{i\phi}}$. The actual expression of this term would be difficult to derive, but we can still infer the effect of $g$ on the final similarity profile. We start by expressing $\overline{g e^{i\phi}}$ as differing from $\overline{e^{i\phi}}$ by a complex number $a e^{ib}$, reading 
\begin{equation} \label{differ}
\overline{g e^{i\phi}}=a e^{ib}\overline{e^{i\phi}}, 
\end{equation}
with $a$ and $b$ unknown dimensionless functions of $kx$. We can be more specific on the behaviour of $a$ and $b$. First, For no displacement ($x=0$), we have $a=1$ and $b=0$, as $\overline{g}=1$. Also, as $kx$ is small in our range of measurement (0.26 for $x$ = HWHM), we can Taylor expand $a$ and $b$ around zero. By keeping only the first non zero terms of the Taylor expansions, we have $a=1+\alpha (kx)^2/2$ and $b=\beta kx$, with $\alpha$ and $\beta$ small dimensionless numbers. Indeed, it can be shown that the derivative of $a$ is zero at $kx=0$ (by expanding the derivative with respect to $x$ of the expression $\overline{g e^{i\phi}}=a e^{ib}e^{i\mu-\sigma^2/2}$). Using these expansions, we can insert (\ref{differ}) in place of $\overline{e^{i\phi}}$ in (\ref{integral}), perform the integral and see the impact this has on the final form. We find the modified profile
\begin{equation} 
S =   \frac{ 1 }
{\big (1-\frac{\sigma^2+\alpha (kx)^2}{2\ln{\rho}} \big)^2+\big (\frac{\mu+\beta kx}{\ln{\rho}} \big)^2}.
\end{equation}

In the axial case, we see that the modification is negligible, as the sigma term was already negligible in the original profile, and $\mu\gg\beta kx$. In the transverse case, we have 
\begin{equation} 
S=  \frac{1}{\left (1-(1+8\alpha)\frac{\left(kx\right)^2}{16\ln{\rho}} \right)^2  +  \left(\frac{\beta kx}{\ln{\rho}} \right)^2},
\end{equation} 

\noindent where we see that the modification is not negligible. This modified profile with two free parameters $\alpha$ and $\beta$ gives an excellent fit of the data for $\alpha=0.027$ and $\beta=0$, which is an argument in favour of the hypothesis that the observed deviation does come from an imperfect Lambertian reflectance and/or a non uniform reflectivity. The modified profile is shown in Fig. \ref{fig:disp}. Note that this modification acts as a scaling of the original profile (\ref{eq:Str}) in the $x$ direction by 10\%.

\section*{Acknowledgements}
This work was supported by funding from the Leverhulme Trust (RPG-2017-197) and the UK Engineering and Physical Sciences Research Council (EP/P030017/1, EP/R004854/1).

\bibliography{sample}



\end{document}